\pdfoutput=1
\documentclass[10pt,journal,cspaper,compsoc]{IEEEtran}
\usepackage[cmex10]{amsmath}
\usepackage{flushend,epsfig,epstopdf,subfigure,booktabs,color,graphics}
\newcommand{\argmin}{\operatornamewithlimits{arg\ min}}
\newcommand{\argmax}{\operatornamewithlimits{arg\ max}}

\begin{document}
\title{Hidden Markov Estimation of Bistatic Range From Cluttered
  Ultra-wideband Impulse Responses}

\author{%
  Merrick McCracken and Neal Patwari%
  \thanks{ M.~McCracken and N.~Patwari are with the Department of
    Electrical and Computer Engineering, University of Utah, Salt Lake
    City, USA. This material is based upon work supported by the
    National Science Foundation under Grant Nos. \#0748206 and
    \#1035565 and by the University of Utah Research
    Foundation. Contact email: merrick.mccracken@gmail.com.}%
}

\markboth{IEEE Transactions on Mobile Computing}{McCracken and
  Patwari: Hidden Markov Est. of Bistatic Range From Cluttered
  UWB Impulse Responses}

\maketitle

\thispagestyle{empty}

\begin{abstract}
  Ultra-wideband (UWB) multistatic radar can be used for target
  detection and tracking in buildings and rooms. Target detection and
  tracking relies on accurate knowledge of the bistatic delay. Noise,
  measurement error, and the problem of dense, overlapping multipath
  signals in the measured UWB channel impulse response (CIR) all
  contribute to make bistatic delay estimation challenging.  It is
  often assumed that a calibration CIR, that is, a measurement from
  when no person is present, is easily subtracted from a newly
  captured CIR.  We show this is often not the case. We propose
  modeling the difference between a current set of CIRs and a set of
  calibration CIRs as a hidden Markov model (HMM). Multiple
  experimental deployments are performed to collect CIR data and test
  the performance of this model and compare its performance to
  existing methods. Our experimental results show an RMSE of 2.85\,ns
  and 2.76\,ns for our HMM-based approach, compared to a thresholding
  method which, if the ideal threshold is known \emph{a priori},
  achieves 3.28\,ns and 4.58\,ns.  By using the Baum-Welch algorithm,
  the HMM-based estimator is shown to be very robust to initial
  parameter settings.  Localization performance is also improved using
  the HMM-based bistatic delay estimates.
\end{abstract}


\section{Introduction}
A useful application of ultra-wideband (UWB) impulse radio is
detection and tracking of people\footnote{In this paper, we use ``people'' or ``person'' to indicate the object being tracked.} in buildings. In particular, bistatic
and multistatic radar systems are used for this application
\cite{Taylor}. This is done by capturing the channel impulse response
(CIR), $h(t)$, between transmitter/receiver pairs and detecting
changes to the CIR.  

This paper describes a contribution to bistatic delay (or equivalently, bistatic range) estimation.  A person induces changes in the CIR starting at the bistatic delay, that is, the earliest time delay at which changes occur in the CIR due to the person being tracked.  If the bistatic delay is denoted $\tau_*$, then the bistatic range is simply the distance this multipath component has traveled, {\it i.e.}, $\tau_* c$ where $c$ is the speed of light.  If RF energy traveled from the transmitter to the person and then to the receiver, with no additional scattering, then the bistatic range defines an ellipse on which the person is located.  Thus bistatic range estimation is a key primitive of UWB tracking systems.

The primary contribution of this work is to develop a method which considers the changes which occur in a CIR at all time delays in order to estimate bistatic delay.  Current published research, as described in Section \ref{S:RelatedWork}, generally are {\it first threshold-crossing} methods, that is, they estimate the bistatic delay as the first delay in which a metric exceeds a threshold.  As a result, they are (a) sensitive to noise in the CIR prior to the true bistatic delay, and (b) sensitive to the correct setting of the threshold parameter.  

Our proposed method uses a hidden Markov model (HMM) to model the changes to the CIR as a function of time delay.  
The Markov chain is a progression between two states: $X=0$, meaning that a person in the environment is not causing changes at the current time delay, or $X=1$, meaning that a person is causing changes at the current time delay. The state of the system is observable only indirectly via the CIR, because of noise and the variability in the multipath channel. The distribution of the observations is dependent on the current state of the system, thus the system is a HMM. Using the observations and the system model, the forward-backward algorithm solves for the most likely state at any given time. The bistatic delay estimate is the time delay at which the system transitions from state 0 to state 1.

When solving for the bistatic delay, our proposed method considers all of the available data and, as we show, the error in bistatic delay estimation is reduced compared to the best thresholding scheme.  Further, using a Baum-Welch algorithm, we avoid the requirement of knowing {\it a priori} the correct parameters.

\subsection{Related Work} \label{S:RelatedWork}

Generally, methods to estimate the bistatic delay or range first perform ``background subtraction''.  This means that a prior measurement, or an average of many prior measurements, of the CIR is subtracted from any current CIR measurement.  These prior measurements are presumed to be made when the area is empty, {\it i.e.}, with a static background.

Some work in UWB-based impulse response radar assumes that background subtraction is completely effective in removing the response due to the static environment \cite{Thoma,Chang,Guohua,Zetik}.  Some work additionally assumes that, after background subtraction, that each single multipath component caused by a person's presence can be distinguished perfectly from the impulses caused by other people and the environment \cite{Chang,Guohua}. In this paper, we show that ranging can still be performed when these assumptions are not true, as is often the case in a cluttered multipath environment.

One way to estimate the bistatic delay is first to perform ``background subtraction'', and then to threshold on the amplitude of the difference.  Zetik et al. \cite{Zetik} describe a thresholding method which uses a simple formula for choosing an appropriate threshold value for accurate range estimation after background subtraction has been performed. Each UWB module has one transmitting and two receiving directional antennas, all relatively close to one another. This makes each UWB module approach a monostatic radar configuration. All of the sensor nodes were pointed inward toward an empty room using directional horn antennas for their experiments.  In contrast, our measurements are performed in furnished office environments, and the additional clutter can make background subtraction less effective.  The estimation methods described in \cite{Zetik} will be used in this work for comparison.

Another way to estimate the delay is to perform a cross correlation of the received signal with a known target scattering profile and then to threshold the correlation values. SangHyun Chang et al.\ approach detection by modeling a human body's scattering as a spectral multipath model and cross correlating this model with the received CIRs \cite{SangHyun1,SangHyun2}. Detection is then performed using an adaptive threshold on the cross correlation. In their work they used a
UWB radio similar to those used in this work but in a monostatic radar
configuration. The human body spectral multipath model was obtained
using empirically collected data from their UWB radio.  They collected
data of a moving human subject in an open field where there was little
or no multipath propagation to validate their detection
method \cite{SangHyun1}. They expanded the method to tracking a human
target and tested it using additional data collected from the UWB
radio \cite{SangHyun2}. The experimental data for tracking was also
collected in an open field.  In contrast, we use measured data from cluttered environments to show that our method is robust to the indoor multipath channel.

Our work is not the first to propose using HMMs for tracking, however, it is the first, to our knowledge, to propose using a HMM for UWB impulse radar bistatic delay estimation.  Nijsure et al.\ used a HMM to model movement in a UWB radar-based tracking system and simulated its performance \cite{Nijsure}.  In their work, the states of the model are non-overlapping geographic regions near the radios rather than changes to the received signal. The measurements in \cite{Nijsure} are unambiguous power delay profiles. In contrast, our HMM is used to estimate the bistatic range, with only two states, whether or not the CIR is impacted by a person at a given time delay or not.  Two-state HMMs have been used in other applications, for example in detecting channel use in dynamic spectrum access \cite{Sun}. The work in \cite{Sun} simulated channel access by primary users and the performance of detection by secondary users, who would use the channel opportunistically, using a HMM-based estimator to detect whether a primary user is currently transmitting. Simulations showed improved detection performance for the HMM-based method compared to a threshold-based method.

\subsection{Organization}

This paper is organized as follows.  %
Section \ref{sec:Methods} describes the methods proposed in this work to estimate $\tau_*$ using hidden Markov models.
Section \ref{sec:data_collection} describes the data collection campaigns carried out to test the proposed methods empirically.
Results for our proposed methods as well as those from performing simple thresholding and the thresholding method described in \cite{Zetik} are reported in Section \ref{sec:results}.
Finally, conclusions are discussed in Section \ref{sec:conclusions}.

\section{Methods} \label{sec:Methods}

\subsection{Measurements}

\begin{figure}[htb]
  \centering
  \subfigure[Static Environment]{
    \includegraphics[width=1.52in]{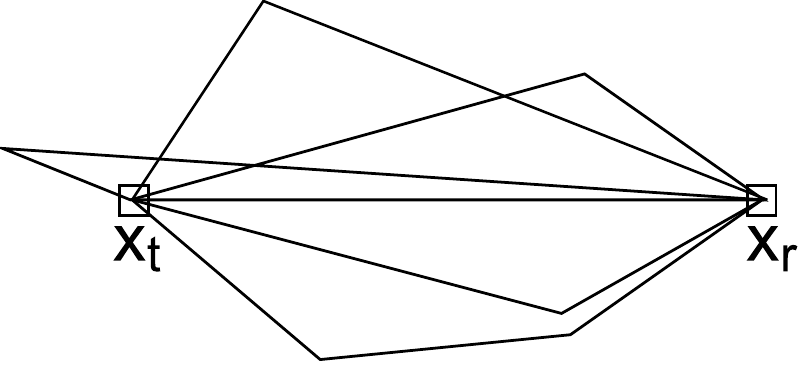}
    \label{fig:without_obj}
  }
  \subfigure[Person's Effect]{
    \epsfig{file=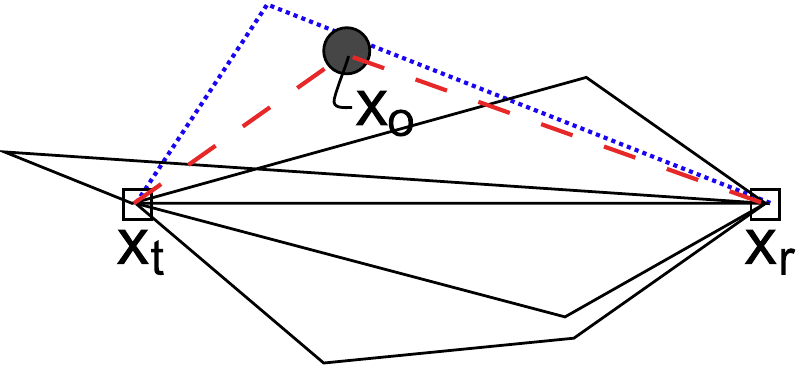,width=1.52in}
    \label{fig:with_obj}
  }
  \caption{When a person appears at $x_0$ in the environment between
    the transmitter at $x_t$ and receiver at $x_r$, he causes an
    additional path with path length $\|x_t-x_0\|+\|x_0-x_r\|$, and
    also affects multipath components with longer path lengths.}
  \label{fig:radiopath}
\end{figure}

Assume that an UWB transmitter sends pulse $\delta(t)$. Due to
multipath propagation, the received signal is described by
\begin{equation} \label{E:h_t}
h(t) = \sum_i \alpha_i \delta(t-\tau_i),
\end{equation}
where $\alpha_i$ and $\tau_i$ are the complex amplitude and time delay
of the $i$th path, respectively. The line of sight path delay is
$\tau_0$. The receiver radio approximately measures the channel
impulse response convolved with the pulse
shape. Fig. \ref{fig:without_obj} is an example of how the transmitted
pulse may follow many different paths to arrive at the receiver.

The number of multipath components seen by the receiver depends on the
environment around the radios. When a person enters the environment,
the person's body will cause a new multipath component at the receiver
as well as affect existing multipath components. This is illustrated
in Fig. \ref{fig:with_obj}.  The delay associated with this new
multipath component is $\tau_*$, which we refer to as the {\it bistatic delay}. 
The person also affects many $\alpha_i$ for $\tau_i \ge \tau_*$.

In bistatic or multistatic radar systems, the bistatic delay,
described by $\tau_*$, is used to locate and track objects near the
radio transmitters and receivers.  Assuming component $i$ is a
single-bounce path ({\it i.e.}, the path is affected by only one scatter as
it travels from transmitter, to the target, and then to the receiver),
the scatter is located on an ellipse with foci at the transmitter and
receiver locations. That is, the locations where the scatter may be
located are points $S$ where the distances from $S$ to the transmitter
and receiver, $S_t$ and $S_r$, sum to:
\begin{equation}
S_t + S_r = c*\tau_i,
\end{equation}
where $c$ is the speed of light.

This work seeks to accurately estimate the bistatic delay $\tau_*$, that of the path
created by the person, particularly in environments with ``cluttered''
impulse responses, {\it i.e.}, those where individual multipath components arrive 
closely in time and become difficult to separate from the CIR.  
Estimation of $\tau_*$ is a key primitive operation for UWB impulse radar systems -- 
estimates from multiple transmitter and receiver pairs can be used to determine possible scatter locations under a single-bounce assumption, as we explore in Section \ref{sec:localization_methods}.

\begin{figure}[htbp]
    \centerline{\epsfig{file=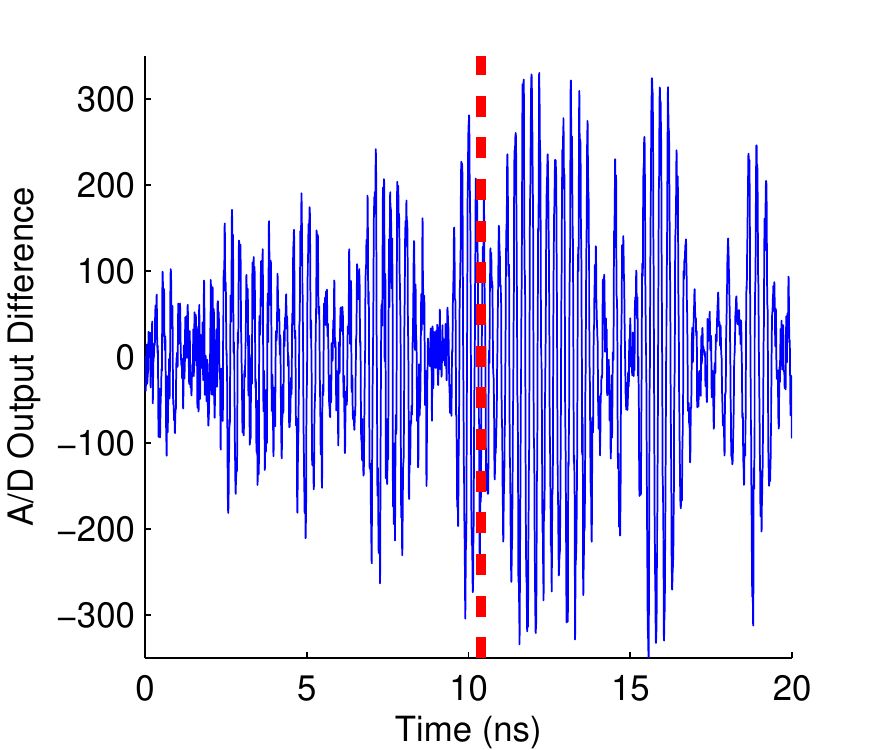} }
	\caption{The difference between a calibration CIR and a new
          CIR gives a noisy signal with multipath components that are
          indistinguishable from one another. The red, dashed line is
          the actual bistatic delay, {\it i.e.}, $\tau_*$.}
	\label{fig:cir_diff}
\end{figure}

As described in Section \ref{S:RelatedWork}, background subtraction is a standard
method for removing the static background CIR from a current CIR measurement.  However,
we have found that background subtraction is not effective in cluttered environments. 
An example is
shown in Figure \ref{fig:cir_diff}, which shows a captured CIR
subtracted from a calibration CIR over about 20\,ns of time, and the true bistatic delay $\tau_*$. Individual
multipath components are indistinguishable and the signal is very noisy.  If background subtraction were effective, the amplitudes prior to $\tau_*$ would be significantly lower than the amplitudes after $\tau_*$, however, this is not the case.  Better methods than simple subtraction to quantify the changes in the CIR are needed.

\subsection{Quantification of Change}

We describe in this section an alternative to background subtraction.  We introduce a divergence measure which quantifies the change between the signal energy measured during the period when the environment is static and the current period. 

We consider a discrete-sampled version of the signal energy, $r_k$,
given by
\begin{equation} \label{E:r_k}
r_k = \int_{(k-1/2)T}^{{(k+1/2)T}} |h(t)|^2 dt,
\end{equation}
where $T$ is the sampling period.  For example, in our experimental
work, we use $T=1$\,ns.  Essentially, $r_k$ is the energy in multipath
components contained within a $T$-duration window near time delay
$kT$. We call this $T$ duration window ``range-bin $k$''. The vector
$\mathbf{r} = [r_1,\ldots, r_n]^T$ is the sequence of $r_k$ samples.
We choose to estimate the energy in each range-bin rather than using
deconvolution to find the CIR. This is done to avoid the problem of
deconvolution generating multiple paths when multipath experience
frequency distortion~\cite{Donlan}.

In this work we use the Kullback-Leibler divergence to quantify the change in the signal energy $r_k$ at each time $k$.
The Kullback-Leibler divergence is a measure of how many additional
bits would be required to encode the samples of one distribution
relative to another distribution. This is also known as relative
entropy \cite{Cover}.

For continuous distributions the asymmetric KL divergence is defined
as
\begin{equation} \label{E:kl_general}
D(p(x)\|q(x)) = \int p(x) \log \frac{p(x)}{q(x)} dx
\end{equation}
where $p(x)$ and $q(x)$ are the probability densities of $r_k$ for the
calibration measurements and for those under test, respectively. The
symmetric KL divergence is defined as $D(p(x)\|q(x)) + D(q(x)\|p(x))$.

The observation signal, $O_k$, in this model represents the difference
between $r_k$ and $r_k$ of the empty room, that is, the calibration
samples. In this work, this difference was calculated as the symmetric
Kullback-Leibler (KL) divergence.

For the observed signal, $O_k$, we use the symmetric KL divergence
assuming Gaussian distributions for $r_k$.  This measure is given in
closed form by,
\begin{equation} \label{E:kld}
O_k = \frac{1}{2}\left(\frac{\sigma_p^2}{\sigma_q^2} +
  \frac{\sigma_q^2}{\sigma_p^2} +
  \frac{\left(\mu_p-\mu_q\right)^2 
  \left(\sigma_p^2 + \sigma_q^2\right)}{
  \sigma_p^2\sigma_q^2}\right) - 1
\end{equation}
where $\mu_p$ and $\sigma_p^2$ are the mean and variance of $r_k$
during calibration, and $\mu_q$ and $ \sigma_q^2$ are the mean and
variance of $r_k$ from the CIR measurements collected for
testing. This closed form solution for $O_k$ is non-negative and the
pdf $f_{O,i}$ will allow us to estimate $X_k$ by applying our hidden
Markov model.

\begin{figure}[htbp]
    \centerline{\epsfig{file=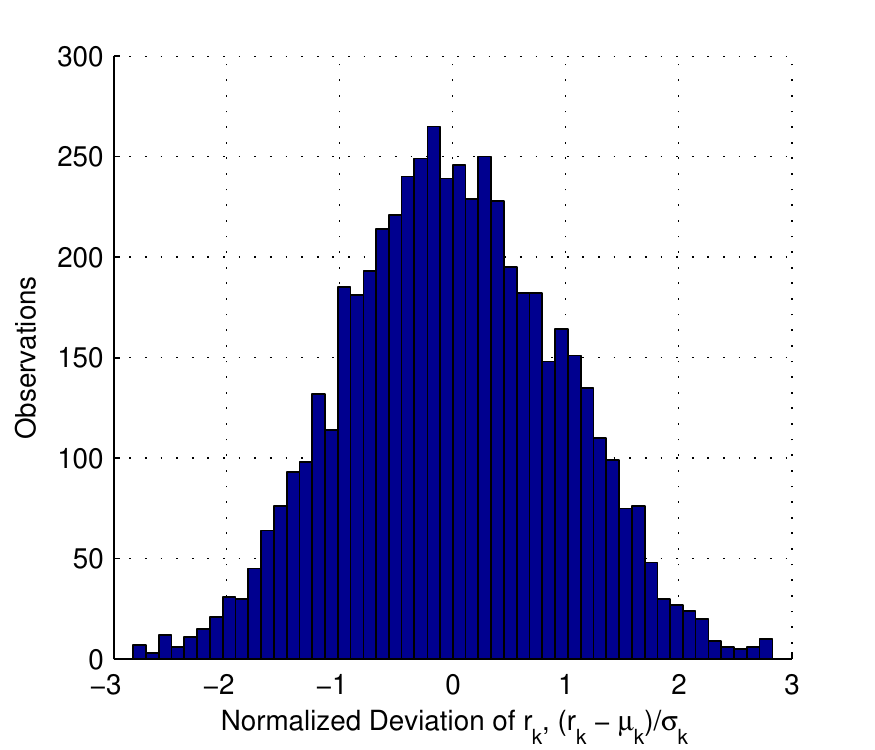} }
	\caption{Empty room samples normalized to have zero mean a
          variance of 1 exhibit a Gaussian distribution.}
	\label{fig:hist_er_normalized}
\end{figure}

The assumption that $r_k$ is Gaussian is important to the closed form
solution of $O_k$ given in equation \ref{E:kld}. To show that $r_k$
follows a Gaussian distribution, each set of 10 samples of $r_k$ for
the empty room was normalized to have a mean of 0 and a variance of
1. These samples were then aggregated for testing. With 10 sets of 90
samples of $r_k$ for the six radio pairs gives 5400 samples. A
histogram of the normalized samples is given in Figure
\ref{fig:hist_er_normalized}. Submitting these samples to a
Kolmogorov-Smirnov test fails to reject the null hypothesis that they
come from a standard normal distribution with $p=0.198$.

\begin{figure}[htbp]
	\centerline{\epsfig{file=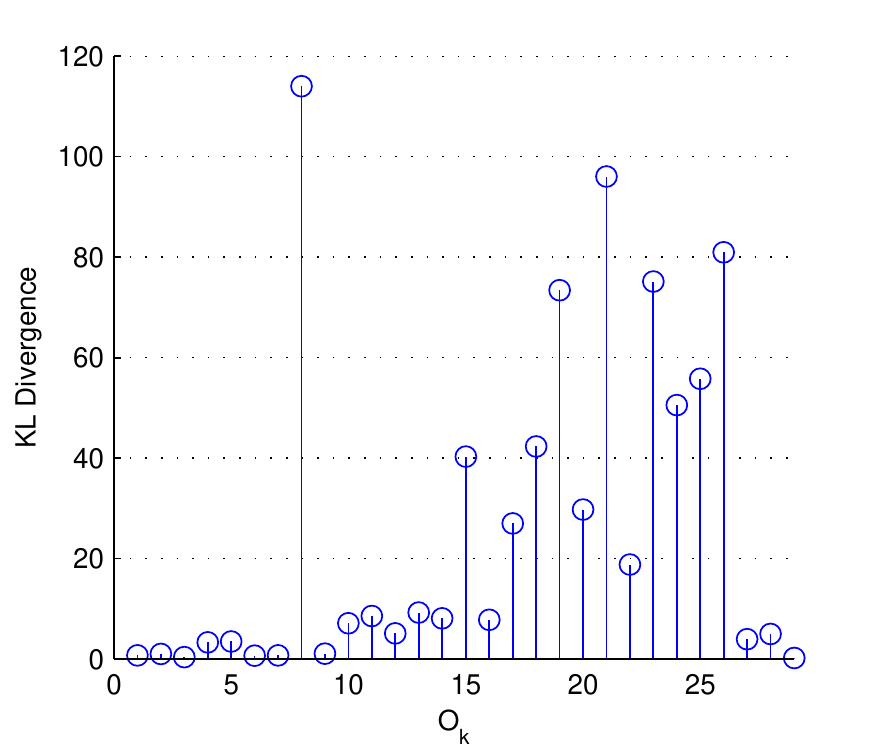} }
	\caption{An example of an observation vector where no
          threshold can find the true $\tau_*$, which is 15 in this
          case. The HMM correctly estimated $\tau_*$ for this vector.}
	\label{fig:hmm_over_ed}
\end{figure}

An example of an observation vector $\mathbf{O}$ of KL divergences is
given in Figure \ref{fig:hmm_over_ed}. This particular example is one
where a first threshold-crossing method would be unable to correctly estimate the
true bistatic delay, $k_*$, of 15. This example shows how the assumption
of easily being able to discern the background signal from the changes
to the CIR can sometimes be wrong. In this case,
there is a very large divergence at a time when the signals should
have shown little or no difference. 

Other distance measures or distributions could be
applied. However, the KL-divergence and Gaussian assumption provide a
standard approach for this proof-of-concept study.

\subsection{CIR Changes as a Hidden Markov Model}

\begin{figure}[hbt]
	\centerline{\epsfig{file=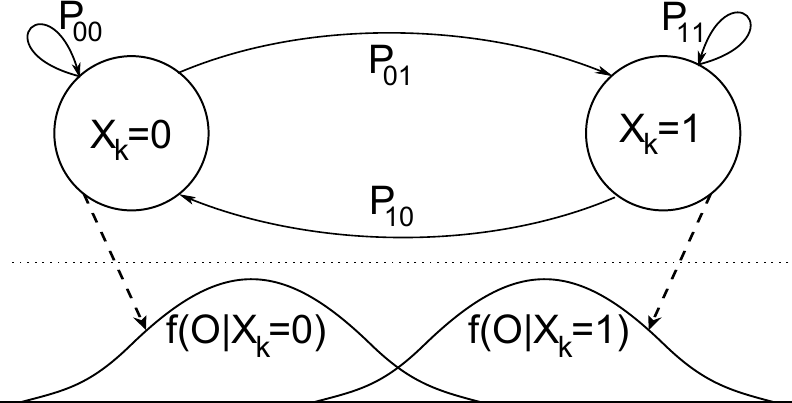} }
	\caption{The change in CIR measurement we observe at range-bin
          $k$, $O_k$, has a distribution dependent on the state,
          $X_k$, of a hidden Markov chain.}
	\label{fig:HMM}
\end{figure}

A hidden Markov model is a special case of a Markov chain.  The
states of a HMM are not directly observable but may be inferred. Other
signals available for observation help determine the past and current
states of the system.  Let $\pi_i$ be the probability of initially
starting the HMM in state $i$, $P_{ij}$ is the probability of
transitioning from state $i$ to state $j$, and $f_{O,i}$ is the
probability of observing signal $O$ given the HMM is in state $i$,
that is, $f(O|X_k=i)$.  A simple illustration of a hidden
Markov model is shown in Figure \ref{fig:HMM}.

In the case when the observations are continuous,
we use the probability density function (pdf) conditioned on the
state, $f_{O,i}$, for a continuous valued random variable. This is
the typical way to describe a HMM for continuous-valued observations
\cite{Rabiner}.

By knowing $f_{O,i}$, $P_{ij}$, and $\pi_i$, a best estimate of the
current state at each time, $\hat{X}_k$, can be calculated. This is
found by applying the forward-backward algorithm to the sequence of
observation signals. When the estimated states transition from
$\hat{X}_k=0$ to $\hat{X}_{k+1}=1$, this gives an estimate for $k_*$
and indicates the presence of a person due to the changes to the
observation vector.

Estimation of $k_*$, where $k_*=\lfloor\frac{\tau_*}{T}\rfloor$, is
equivalent to estimating $\tau_*$. Due to multipath scattering and the person's
impact on those later-arriving signals, $r_k$ will experience
changes, or $X_k = 1$, for many $k\ge k_*$.  The advantage of applying
a HMM is that information over all $k$ is considered when solving for
$X_k$ rather than considering values at each $k$ independently of
changes at all other $k$.

A more thorough introduction to
hidden Markov models and the algorithms used to infer information
about them can be found in \cite{Rabiner}.

\subsection{Continuous Observation Densities}
The observations $O_k$ are continuous valued and their probability
distribution is described by $f_{O,i}$, the probability
density function of $O_k$ given $X_k=i$, $i\in \{0,1\}$.
The HMM parameters $f_{O,i}$, $\pi_i$, and
$P_{ij}$ are estimated using the data $\mathcal{D}$ collected in one
room and are used as initial estimates of the HMM parameters when
estimating $k_*$ for the other room.

The data sets $\mathcal{D}_{i}$, for each state $i$, are made using the
knowledge of $k_*$ by
\begin{equation} \label{E:d_0}
\mathcal{D}_{0} = \{O_k | k < k_* \}
\end{equation}
\begin{equation} \label{E:d_1}
\mathcal{D}_{1} = \{O_k | k \ge k_* \}
\end{equation}

Dividing the observation signals in this way assumes that there will
only be one transition from state 0 to state 1 and no transitions back
to state 0, that is $P_{10} = 0$ and $P_{11} = 1$.

Under the assumption that $X_k = 1$ given $k \ge k_*$, one may also
assume that $P_{1,0} = 0$ and $P_{1,1} = 1$, that is, $P(O_k|X_k=1)$
remains constant as $k$ increases.  This assumption may not be true -- a
person's effect will eventually diminish for large $k$.
Also, a probability of $0$ leaves little opportunity for change
during optimization. To improve the model, we allow a small
probability of returning from state 1 to state 0, i.e., set
$P_{10}=\epsilon$ where $\epsilon$ is a small value greater than 0.

\begin{figure}[hbt]
	\centerline{\epsfig{file=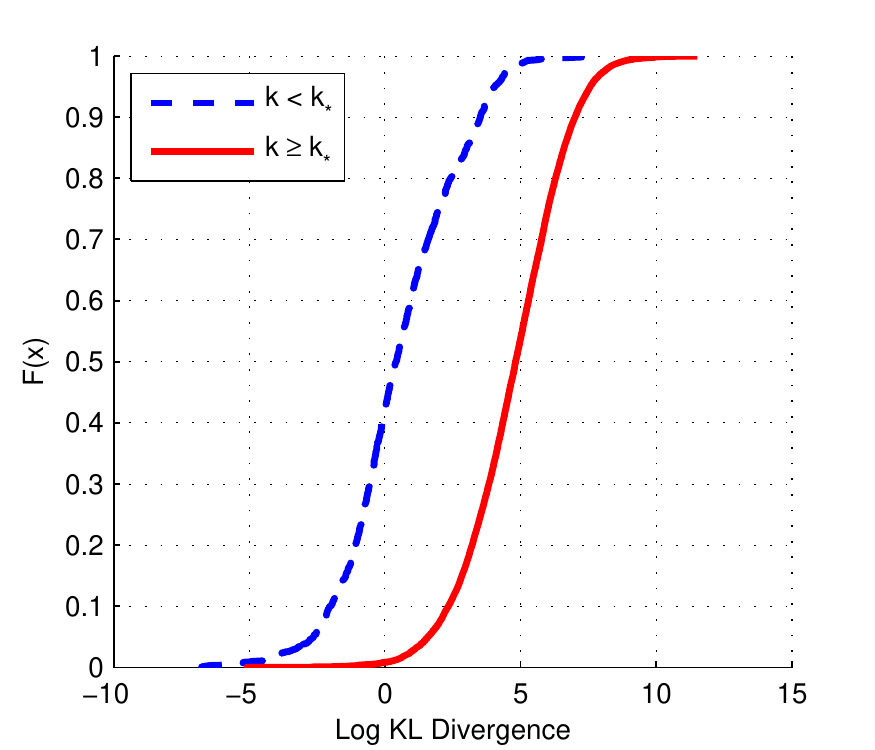} }
	\caption{Empirical CDFs of the log of $O_k$ for one
          room. Although these 
          distributions are not precisely log-normal, this assumption
          is reasonable for the solving methods.}
	\label{fig:log-normal}
\end{figure}

In \cite{McCracken}, no assumptions were made regarding the
distribution the observations took on. The distribution was estimated
by performing an Expectation Maximization algorithm to fit the
data to a Gaussian mixture model. This operation was computationally
expensive but effective. In this work we utilize our observation that the
densities are similar to a log-normal distribution. Under this
assumption, well known maximum-likelihood estimates are used for the
distribution parameters. Figure \ref{fig:log-normal} shows the
empirical CDFs of the aggregate samples before and after $k_*$ for one
room. The natural log is applied to $O_k$ in these distributions. This
log-normal approximation reduces the computational load without
sacrificing solving accuracy.

Initial estimates for $\pi_i$ and $P_{ij}$ are given by
\cite[eq. (40a-b)]{Rabiner} using the training data.

\subsection{HMM Solving} \label{sec:hmm_method}
The HMM parameters are described by $\lambda$ as
\begin{equation} \label{E:lambda}
\lambda = \left[ \pi_i, P_{ij}, f_{O,i} \right]
\end{equation}

The data from one room is used as training data to obtain an initial
estimate of $\lambda$ to begin solving for $k_*$ with the other room's
data, or that of the measurement room.  The following describes how
$k_*$ is estimated for the measurement room once $\lambda$ is
estimated from the training data, as described previously.

Finding $\hat{X}_k$, the estimate of $X_k$, for the measurement room
is done by solving the forward-backward algorithm.
This algorithm finds the most likely state $X$ at each range-bin $k$
\cite{Rabiner}. 
\begin{equation}
\hat{X}_k = \argmax_i P(X_k=i|\mathbf{O},\lambda)
\end{equation}
The forward-backward algorithm is different than the Viterbi algorithm, which finds the most
likely state sequence over all $k$. It may seem more appropriate to
use the Viterbi algorithm to estimate when the state change
occurs. The Viterbi algorithm, however, only returns a state
sequence. By using the forward-backward algorithm, the additional
uncertainty information of $P(X_k=i|\mathbf{O},\lambda)$ is available
for each $k$ when performing localization.  It should be noted that
estimates for $k_*$ are not constrained by the room boundaries or any prior information about where the person might be located.

After estimates for $X_k$ are obtained, the Baum-Welch algorithm uses
these estimates to update the set of HMM parameters such that
\begin{equation} \label{E:update_lambda}
P(\mathbf{O}|i,\lambda_{n+1}) > P(\mathbf{O}|i,\lambda_{n})
\end{equation}
This is algorithm an iterative optimization on the space of
$\lambda$ to maximize $P(\mathbf{O}|i,\lambda_{n})$.

The HMM parameters are updated over all sets of $\mathcal{D}$ as
described by Rabiner~\cite{Rabiner}.
Also, $f_{O,i}$ is again found by estimating the distribution
as log-normal using $\mathcal{D}_i$. However,
$\mathcal{D}_i$ is now found as
\begin{equation} \label{E:d_i}
\mathcal{D}_{i} = \{O_k | \hat{X}_k = i\}
\end{equation}

The algorithm continues for a predetermined number of
iterations or until $P(\mathbf{O}|\lambda_{n})$ no
longer increases more than a given tolerance with each iteration, that
is, $P(\mathbf{O}|\lambda_{n}) - P(\mathbf{O}|\lambda_{n-1}) < \epsilon$.
The final estimate for $k_*$ is
\begin{equation} \label{E:k_*}
\hat{k}_*^{HMM} = \argmin_k P(X_k=1|\mathbf{O},\lambda) > 0.5
\end{equation}

This finds a local maximum in the space of possible $\lambda$ but may not find the global maximum.  The effectiveness of
this algorithm is dependent on the initial values of the HMM
parameters and the data itself. Other optimization algorithms exist
but were not explored in this research.

\subsection{First Threshold Crossing} \label{sec:thesholding}

A standard method to determine the bistatic delay, $k_*$, is
simply to find the first time at which $O_k$ is greater than a
threshold. We refer to this method as {\it first threshold
  crossing} (FTC). Specifically the estimate of $k_*$ in first threshold crossing is given by
\begin{equation} \label{E:energy_detection}
\hat{k}_*^{FTC} = \argmin_k O_k > \gamma
\end{equation}
where $\gamma$ is a threshold. We show the performance of this method
in Figure \ref{fig:energy} as a function of $\gamma$.
To show how the method would perform with training, we assume that
$\gamma$ is set by using the $\gamma$ that achieves the lowest room
mean squared error (RMSE) in one room, and test performance with that
$\gamma$ in the other room. 

The work presented by Zetik et al.\ in \cite{Zetik} gives another
method for thresholding the received CIR to estimate $\tau_*$.  This
method is also used for comparison in Section
\ref{sec:threshold_results}.

\subsection{Localization} \label{sec:localization_methods}
Multiple range estimates allow localization to be performed.  In this section, 
we describe methods for merging bistatic range estimates to obtain a position
estimate.  Clearly, range estimates contain errors, and any location
estimator must deal with these noisy inputs.

One advantage of the HMM-based approach we propose in this paper is that 
it provides a "soft" decision on the bistatic range estimate.  The forward-backward algorithm
quantifies the probability of each state $i$ at each time index $k$, 
$P(X_k=i|\mathbf{O},\lambda)$.  If the conditional probability of state 1 
increases from zero to one very quickly at time $k$, the data is very clear that the delay bin $k$
is very likely to have been the bistatic delay.  If the conditional probability increases
slowly from zero to one over several delay bins, then the data is less clear.  Essentially,
a quantification of the probability of each delay bin $k$ being the bistatic delay is
given by the rate at which the conditional probability changes.

The forward-backward algorithm finds the conditional probability of being in
a given state at time $k$. To simplify notation going forward, we will
let $\alpha_k = P(X_k=1|\mathbf{O},\lambda)$. Since there are only
two states, $\alpha_k$ fully describes the probability of being in a
given state at time $k$. Also, let $(x)^{+}$ be defined by
\begin{equation}
(x)^{+} = 
   \begin{cases}
     x & \text{if } x \geq 0 \\
     0 & \text{if } x < 0
   \end{cases}
\end{equation}
Assuming a single-bounce model, each time delay measurement corresponds to a region on the plane given
an ellipsoid with the transmitting and receiving radios at the
foci. For a location estimate on a 2D plane, at least three radio
pairs must give range estimates for the overlapping elliptical regions
to produce a unique solution, assuming noise-free range
estimates. Due to the cluttered environment, whose background UWB
reflections are often much stronger than the ones caused by a person,
the range estimates cannot be assumed to be noise-free. For this work, to mitigate the effect of
having range estimate inaccuracy, we
obtained data from six radio pairs.

Localization can be solved as an inverse problem, described by Cheng Chang
et al.\ as a semi-linear algorithm (SLA) \cite{Chang}  which models the
radio locations and range estimates as a linear function $g=Az$
\cite[eq. (4)]{Chang}. SLA is solved using a linear
least squares method. Where range estimates alone are available,
solving the problem as an inverse problem makes the most sense since
these estimates will often not converge perfectly due to errors and
noise. 

The output of the HMM, however, is more than a simple range
estimate. Additional information about the probability of being in one
of the two HMM states is available. This additional uncertainty at
each time $k$ can be used to improve localization accuracy.

In this work, localization is solved as a forward problem as follows.  We discretize space into $P$ pixels containing the area being monitored.
We denote $l_i$ to be a quantification of  the ``presence'' of a person in pixel $i$.  The image vector is then
\begin{equation}
L = [l_1, \ldots, l_P]^T,
\end{equation}
where pixel $i$ is centered at coordinate $z_i=(x_i,y_i)$. A person in pixel $i$ would, assuming the single-bounce model, be measured to be in range-bin $k^m_i$ for transmitter/receiver pair $m$, where $m \in \{1, \ldots, M\}$,
\begin{equation}
k^m_i = \left \lceil \frac{\|t_m-z_i\| + \|z_i-r_m\| - \|t_m-r_m\|}{
    d_k} \right \rceil
\end{equation}
where $t_m$ and $r_m$ are the transmitter and receiver coordinates for link $m$ and
$d_k$ is the distance light travels during one time bin. The value $l_i$
is given by
\begin{equation} \label{eq:pnorm}
l_i = \left[\sum_{m=1}^M \left[A^m\right]_i^p \right]^{\frac{1}{p}}
\end{equation}
where $A$ is the non-negative difference function of $\alpha$ at $k^m_i$, 
\begin{equation} \label{eq:A}
\left[A^m\right]_i = (\alpha_{k^m_i} - \alpha_{k^m_i -1})^{+}
\end{equation}
with $\alpha_0=0$.  Equation (\ref{eq:pnorm}) is the $p$-norm of $\{A^m\}$ for all radio
pairs $m=1, \ldots, M$ at pixel $i$. A $p$-norm of 0, i.e. $p=0$,
gives a count of non-zero values and a $p$-norm of 1 is a sum of the
elements. In this work, $p=0.2$ was found to
give the best performance and was the value used for the results given
in Section \ref{sec:HMMResults}. This $p$-value weights the elements
of $A$ such that, qualitatively, lower values are weighted more and
higher values are weighted less.

Rather than using $\alpha_k$, localization can also be done using
estimates $\hat{k}_*$. This would change the way $A$ is calculated
from what is given in Equation (\ref{eq:A}) to:
\begin{equation}
\left[A^m\right]_i =
   \begin{cases}
     1 & \text{if } i = \hat{k}_* \\
     0 & \text{if } otherwise
   \end{cases}
\end{equation}
Results for both of these methods for solving localization as a
forward problem as well as solving using SLA are given in Section
\ref{sec:localization_results}.

To understand pixel value $l_i$ more intuitively, we recall that $\alpha_{k^m_i} - \alpha_{k^m_i -1}$ is a soft metric for the probability that pixel $i$ is at the same bistatic range as the person, as indicated by the measurement on link $m$.  Due to the $p$-norm in (\ref{eq:pnorm}), $l_i$ is a type of average of these probabilities over all links.  This method is especially useful when the measurements from a link are ambiguous, and thus $\alpha_k$ for that link doesn't change from zero to one suddenly. The uncertainty in $\{\alpha_k\}_k$ is reflected
in the presence image $L$.

For purposes of noise reduction, we apply a 2-D Gaussian filter to image $L$.  For experiments with one person in the area, we take the coordinate of the pixel with highest $l_i$ (after the filtering) as the location of the person.

\section{Experiment} \label{sec:data_collection}

We conduct two types of experiments for evaluation of our proposed algorithms.  First, we conduct in-room experiments where transmitters and receivers are in the same room as the person being located.  Second, we conduct an experiment in which the transmitter and receiver are on the other side of an interior wall of the room in which the person is located.  In all experiments, we use two P220 UWB impulse radios from Time Domain, Inc., to capture CIR measurements.

\subsection{In-Room Experiments} \label{sec:in_room_method}
\begin{figure}[htb]
  \centering
  \subfigure[Room 3325]{
    \epsfig{file=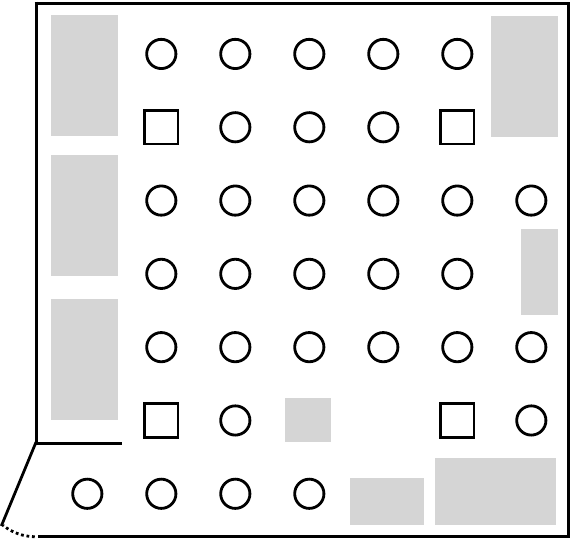}
    \label{fig:Room1}
  }
  \subfigure[Room 1280]{
    \epsfig{file=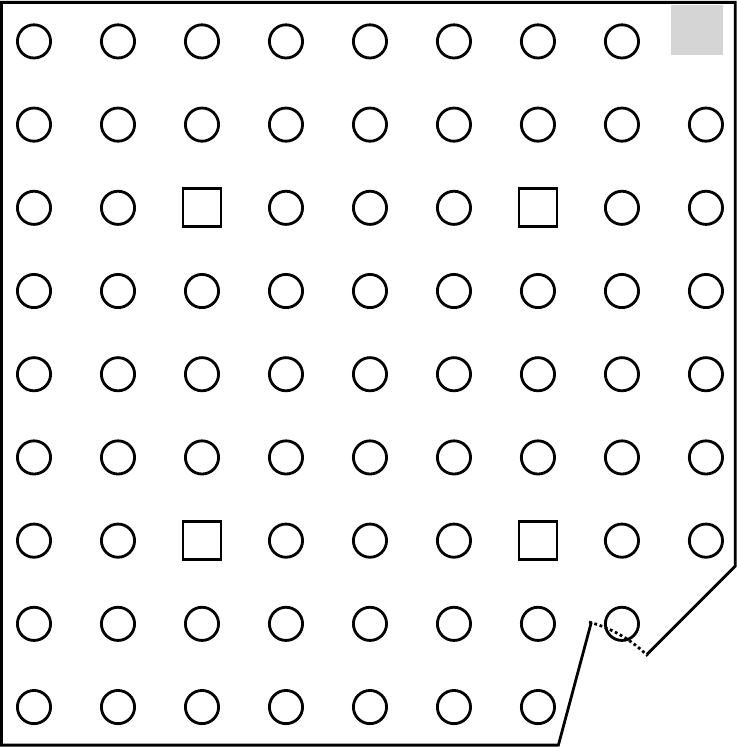}
    \label{fig:Room2}
  }
  \caption{Circles are points where the person would stand and squares
    are radio locations. Gray rectangles are furniture. Neighboring
    points are spaced 90 cm apart.}
  \label{fig:rooms1and2}
\end{figure}

We first conduct measurements in rooms 3325 and 1280 in the Merrill Engineering Building. Two rooms are measured
so that one room can be used as a training room while the other is used as an experiment
room. Figures \ref{fig:Room1} and \ref{fig:Room2} describe the
positions of the radios and where the person stands in each room.
Room 3325 contains typical office furniture; desks, chairs,
bookshelves, and computers. Room 1280 is a classroom and all of the
desks and furnishings were removed from the room for the
experiment. Room 1280 is also larger than room 3325, as shown in
Figure \ref{fig:rooms1and2}.

We collect both empty-room ({\it i.e.}, no person in the room) calibration measurements and measurements which
represent all measurements possible in a four UWB transceiver
multistatic network when a person is standing at any of the possible
grid points in the two rooms. Since we have only two UWB
transceivers, we conduct these measurements as follows.

The two radios are placed in any of the four locations designated for
the radios in the room.  Ten calibration measurements of $r_k$ are
taken when the room is empty. Then, at each of the designated points,
a person stands and remains as motionless as possible while ten more
measurements of $r_k$ are taken.  After collecting measurements at 
all points, the two radios are moved.  This process is repeated for the $M=6$
pair-wise radio locations.  Then, the full process is repeated in the
second room.

Experiment A uses the data collected in room 1208 as the training room
data and the data collected in room 3325 as the data for the
experiment room. Experiment B swaps the data used for the training and
experiment rooms and performs the estimation again.

\subsection{Through Wall Experiment} \label{sec:through_wall_method}

In addition to ranging and localizing a person that is in the same
room as the radios, one data set is also collected to test ranging
through an interior wall. Two radios are placed 1 m apart from one
another and 18 cm from the wall in room 3220 in the Merrill
Engineering Building at the University of Utah. 

We also report the power loss due to wall penetration, in order to 
characterize the experiment condition. To estimate the penetration 
loss of the wall, the CIR is measured with the radios 4.5\,m
apart with both radios in room 3220. The transmitting radio is then
placed on the other side of the wall in room 3230 and the receiving
radio is also moved to maintain a 4.5\,m separation. The CIR is
measured again and the line-of-sight component of two measured CIRs
are compared. The measured power loss of the wall is approximately 5
dB over the 3-5 GHz band.

\begin{figure}[thb]
    \centerline{\epsfig{file=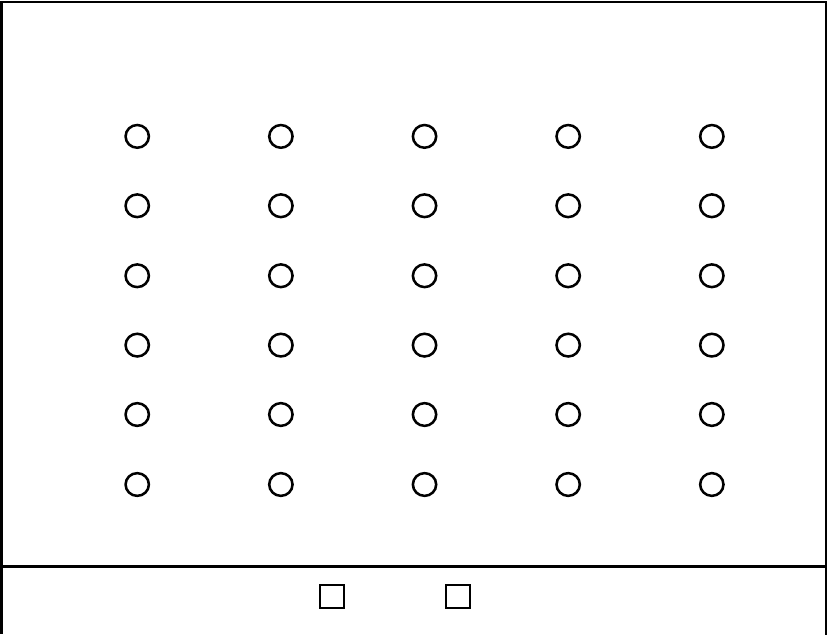} }
	\caption{Squares represent radio locations in room 3220 and
          circles represent person locations in room 3230. Person
          locations are spaced 60 and 120 cm apart.}
	\label{fig:meb3230}
\end{figure}

The measurements are made as follows.  A person stands at 30 different
locations in the adjacent room 3230 while the CIR was captured 20
times per location. Figure \ref{fig:meb3230} shows these two rooms
with their corresponding person and radio locations. Both before and
after all of these CIRs are sampled with a person present, the CIR for
the empty room is captured 100 times. UWB pulse integration is also
increased by a factor of 8 from what was used in the other
experiments. This increases the SNR of each CIR at the cost of
lowering the maximum possible sampling rate.

This through wall experiment is performed for just one radio pair, which 
is insufficient for localization. Instead, the purpose of this through-wall experiment is to allow us to quantify the performance of
UWB impulse radio bistatic delay estimation. 

\section{Results} \label{sec:results}

In this section, we apply the methods proposed in Section \ref{sec:Methods} to the
data collected as described in Section \ref{sec:data_collection}.  We
measure the performance of our proposed HMM-based bistatic delay
estimator in three ways: (1) the RMSE of the bistatic delay estimator,
(2) the false negative and false positive rates, and (3) the
performance of localization using our bistatic delay estimates. We
compare the results of our method of estimating bistatic delay to
simple thresholding as well as the thresholding method given in
\cite{Zetik}.

The bistatic delay error is the difference between the person's actual bistatic delay and the estimated bistatic delay, 
\[
  \varepsilon = T \left| \hat{k}_* - k_* \right| 
\]
We use root mean-squared error (RMSE) across all experiments to quantify average performance.

We report false negative and false positive rates for the methods
studied.  For bistatic delay estimation, a false negative is when
there was no person's bistatic delay detected when a person is
actually present.  For our HMM-based method, this corresponds to the
forward-backward algorithm detecting no transition from state 0 to
state 1 for the measured CIR. A false positive is when there was a
bistatic delay is estimated when no person was present.

In all results, we chose a delay-bin duration $T$ of 1\,ns. The
choice of $T$ is a trade-off between computational requirements and
quantization noise. We note that 1\,ns of time corresponds to about
30 cm of distance traveled at the speed of light, approximately the
width of an adult human body.  Further, our results show errors
significantly higher than 1 ns, and thus it has not been necessary for
us to reduce $T$ further.

\subsection{First Threshold Crossing} \label{sec:threshold_results}

\begin{figure}[htbp]
	\centerline{\epsfig{file=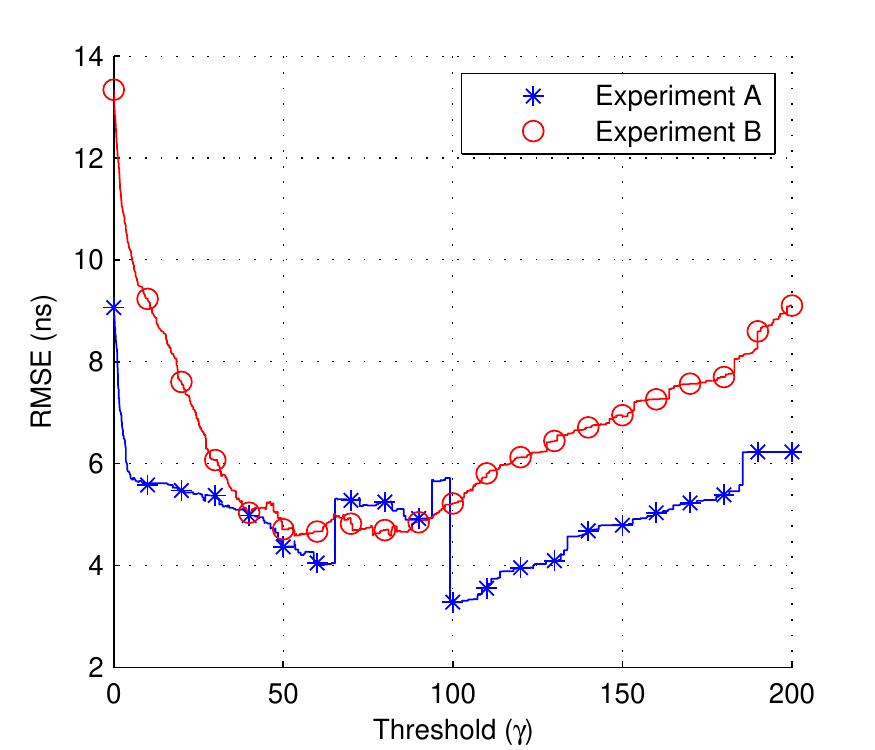} }
	\caption{Performance of {\it first threshold crossing} method
          given by equation
          \eqref{E:energy_detection} as a function of threshold
          $\gamma$.}
	\label{fig:energy}
\end{figure}

\begin{figure}[htbp]
	\centerline{\epsfig{file=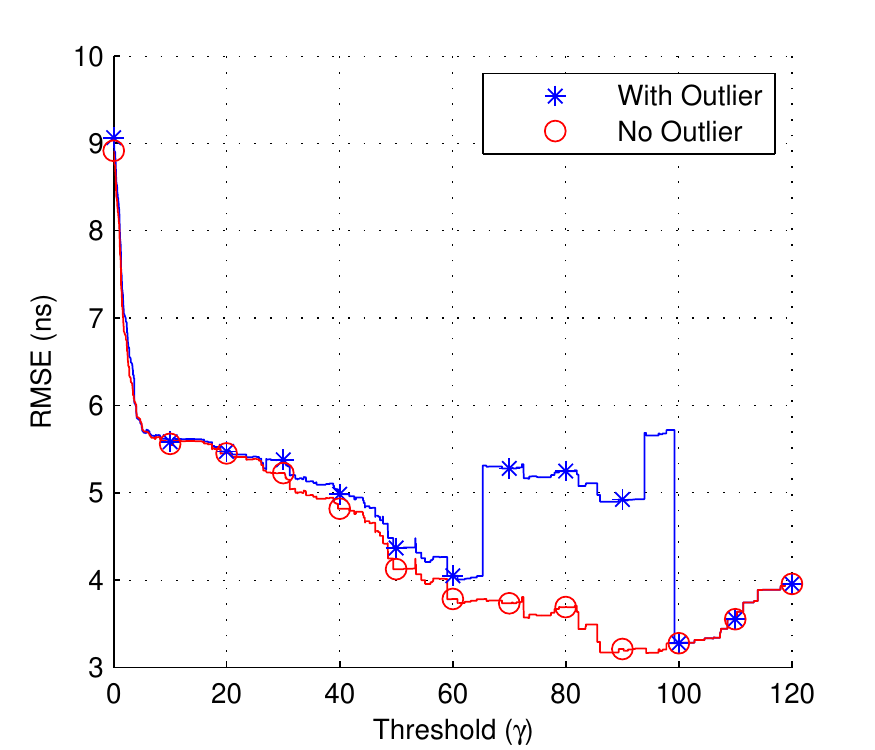} }
	\caption{RMSE for Experiment A with and without the outlier
          point.}
	\label{fig:energy_outlier}
\end{figure}

First, we test the performance of the FTC estimator as described in
Section \ref{sec:thesholding}.  We find the threshold that is
optimal (for minimum RMSE) for the training room and then use that
threshold in the testing room.  From this method, a minimum RMSE of
$5.25$\,ns is achieved for Experiment A and $5.20$\,ns for Experiment
B. Next, we see what minimum could have been obtained for the testing
room even if the optimal threshold for that room had been known.
These absolute minimums achieved are $3.28$\,ns and $4.58$\,ns,
respectively. Figure \ref{fig:energy} shows how the RMSE varies as a
function of the threshold.  Clearly, the optimal threshold would not
be known {\it a priori} for each room. Figure \ref{fig:energy} shows
the sensitivity of the RMSE to chosen threshold.  For Experiment A
there is a large change in the estimates with a small change to
$\gamma$. This large change to the RMSE, occurring near $\gamma$
values of 65 and 99, are due primarily to one set of CIRs for one
point and radio pair. Without knowledge of the true values for $k_*$,
one would still notice the large change to $\hat{k}_*$ with small
changes to $\gamma$. The effect on RMSE due to this one outlier is
shown in Figure \ref{fig:energy_outlier}.

There were no false negatives for the range of $\gamma$ tested in Figure
\ref{fig:energy} for either experiment using the first threshold
crossing method.

The work done by Zetik et al.\ \cite{Zetik} gives a somewhat different
method for thresholding the signals. The background is continually
updated for each UWB node, which would correspond to a radio pair in
our work, as:
\begin{equation} \label{eq:zetik_background}
\mathbf{b}^i = \alpha\mathbf{b}^{i-1}+(1-\alpha)\mathbf{m}^i
\end{equation}
where $\mathbf{b}$ is the background estimate and $\mathbf{m}$ is the
newly measured CIR. The signal $\mathbf{s}$ then used for thresholding
is:
\begin{equation} \label{eq:zetik_background_subtraction}
\mathbf{s}^i = \mathbf{m}^{i}-\mathbf{b}^i
\end{equation}
This removes the static background signal from the time-varying
signal, which is what we wish to detect and range.

The threshold is calculated as: 
\begin{equation} \label{eq:zetik_threshold}
t^i =
\left(0.3+0.7\frac{n^i}{\left|\left|\mathbf{s}^i\right|\right|_{\infty}}\right)
\left|\left|\mathbf{s}^i\right|\right|_{\infty},
\end{equation}
where $n^i$ is the peak noise level of $m^i$.

Using the method of \cite{Zetik}, described in Equations
(\ref{eq:zetik_background_subtraction}), (\ref{eq:zetik_background}), and (\ref{eq:zetik_threshold}),
and the data collected, we obtain an RMSE of $6.5$\,ns and $10.6$\,ns
for experiments A and B, respectively.

\begin{figure}[hbt]
	\centerline{\epsfig{file=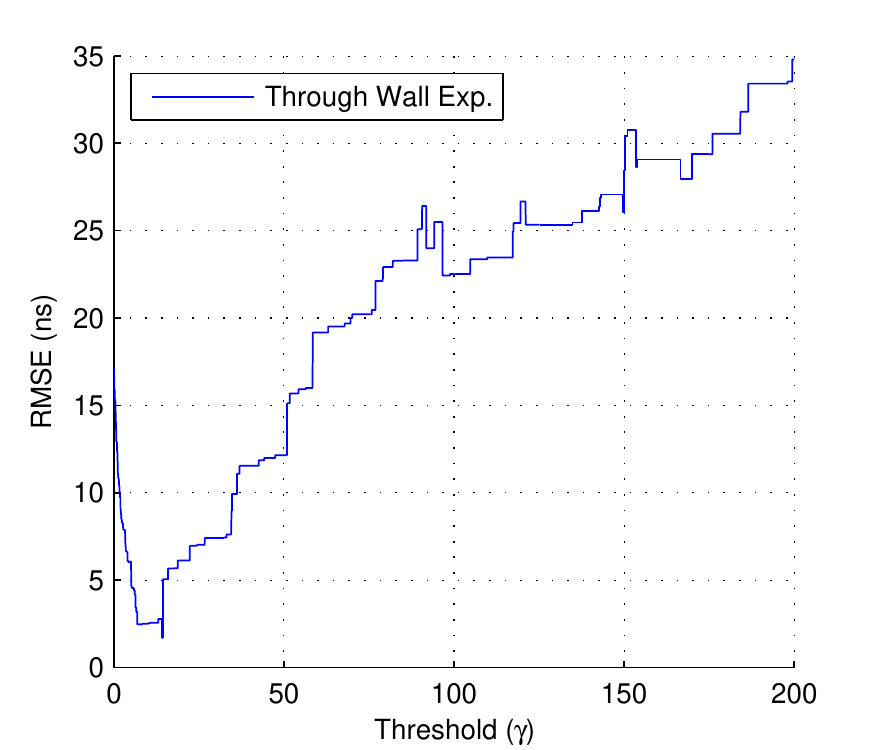} }
	\caption{Performance of {\it first threshold crossing} method
          for the through-wall experiment}
	\label{fig:energy_tw}
\end{figure}

When first threshold crossing is performed on the through-wall experiment
data, a plot of RMSE versus threshold is obtained and 
shown in Figure \ref{fig:energy_tw}.
This is comparable to those shown in Figure \ref{fig:energy}. Notice
that the 
$\gamma$ that achieves the optimal estimation of $k_*$ is different
for each experiment and varies significantly. In other words, the optimal
$\gamma$ cannot be determined from data measured in a different location. 

\subsection{HMM-based Method} \label{sec:HMMResults} 
\begin{figure}[htbp]
	\centerline{\epsfig{file=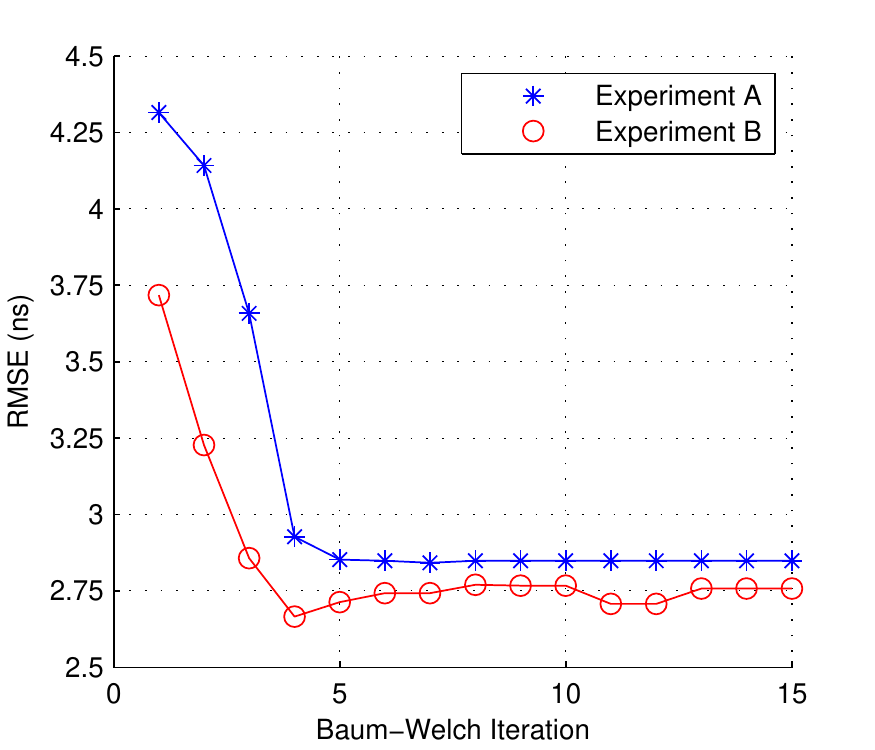} }
	\caption{Performance of HMM-based estimator of $k_*$ as a
          function of iteration count.}
	\label{fig:hmm_rmse}
\end{figure}

The HMM and process described in Section \ref{sec:hmm_method} are
applied to the two in-room experimental data sets. The changes to RMSE for each iteration of the
Baum-Welch algorithm is shown in Figure \ref{fig:hmm_rmse}. The RMSE
achieved after 15 iterations is $2.85$\,ns and $2.76$\,ns for
Experiments A and B, respectively. There were no false negatives. The
bias, $E[\hat{k}_*-k_*]$, was $-0.3$\,ns for Experiment A and
$0.2$\,ns for Experiment B.

The marked improvement in RMSE from using a HMM over energy detection
also comes without foreknowledge of an ideal threshold value. Although
an initial estimate for $\lambda$ is required, the Baum-Welch
algorithm eliminates much of the error due to a poor estimate, as will
be shown with the through-wall results \ref{sec:through_wall}. The
HMM, unlike a simple threshold, takes into account the data across all
time values to estimate $k_*$.

The stopping condition used for the given results is to continue the
Baum-Welch algorithm until there is little change to
$P(\mathbf{O}|\lambda)$ from one iteration to the next. That is
$P(\mathbf{O}|\lambda_{n}) - P(\mathbf{O}|\lambda_{n-1}) < \epsilon$.
Experiment A converges, using this metric, after 9 iterations and
Experiment B after 14 iterations.

\subsection{Through-wall experiment} \label{sec:through_wall} 

Our proposed HMM method is also applied to data captured through a
wall dividing two rooms as described in Section
\ref{sec:through_wall_method}. Observation vectors are calculated
using all of the available empty room CIRs and CIRs with a person
present. With the observation vectors and an initial estimate for the
HMM parameters $\lambda$, estimates for $k_*$ can be found.

Using the $\lambda$ that is found to be optimum for any one of the
three environments as the initial $\lambda$ for any of the other
environments results in the same solution for $\lambda$ from the
Baum-Welch algorithm. This is illustrated using the through wall data.
For the through wall data, there are three choices of $\lambda$, two
obtained from the data collected from the two in-room experiments
described in Section \ref{sec:in_room_method} and one from the data
and known locations of this through wall data. The $\lambda$ obtained
from the through wall data could not be used in a production system
because it is derived using a knowledge of $k_*$. If $k_*$ is known,
there is no reason to use it to find $\lambda$ to then estimate
$k_*$. It is used here solely for illustrative purposes.

\begin{figure}[htbp]
	\centerline{\epsfig{file=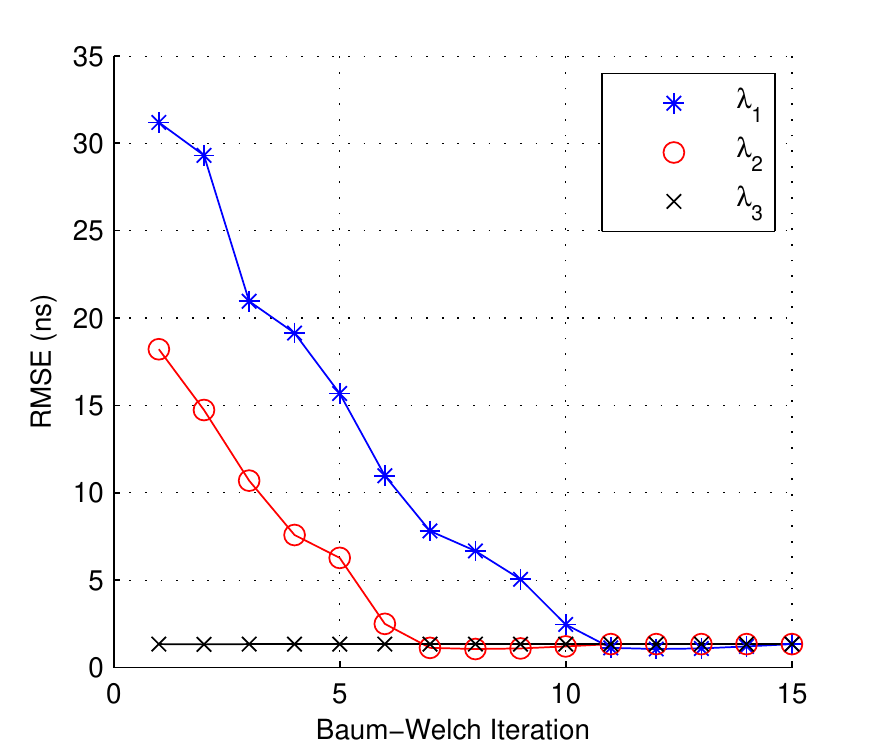} }
	\caption{The RMSE for the through-wall experiment converges to
          1.33\,ns for each of the initial choices of $\lambda$ derived
          from the data for each of the three rooms.}
	\label{fig:baum_tw}
\end{figure}

Figure \ref{fig:baum_tw} shows the bistatic delay RMSE at each
iteration of the Baum-Welch algorithm for the three different choices
for $\lambda$ at the first iteration. The choice of $\lambda$ greatly
influences the RMSE at first, but the effect of the choice is
ultimately negated by the Baum-Welch algorithm.  The final RMSE in all
three cases is 1.33\,ns. 

This final error is better than the results obtained with the subject
in the same room as the radios. There are several reasons for
this. 
\begin{enumerate}
\item {\it Number of samples}: Many more samples of the empty room were collected and
used in determining the KL-divergences in the through-wall experiment (200) compared to the in-room experiments (10). These additional samples help 
to reduce the noise in the observation vectors. The effect of
choosing different empty room samples is explored
further below.  
\item {\it Additional integration}: Additional signal integration was done in
sampling to reduce noise in the CIRs because of the additional path
loss in the through-wall experiment. 
\end{enumerate}

\begin{figure}[htbp]
	\centerline{\epsfig{file=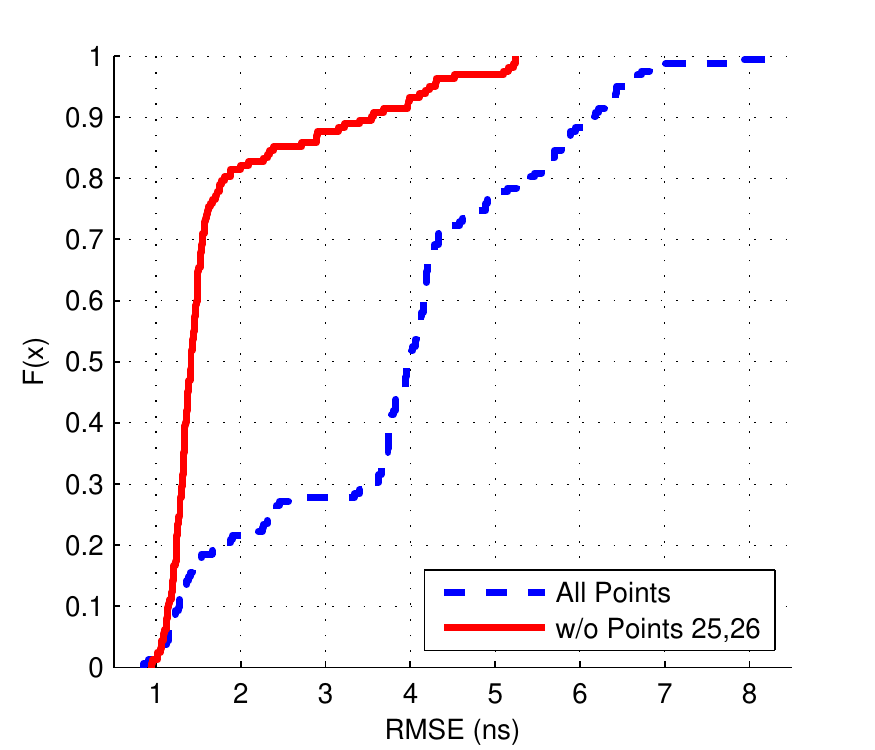} }
	\caption{Variance of the estimator based on which set of empty
        room samples is used.}
	\label{fig:er_ecdf}
\end{figure}

To show the effect of the number of empty room samples on the
performance of the ranging estimation (item 1 above), we run an
experiment in which we reduce the number of empty-room samples used in
the through-wall experiment.  Here, we calculate observation vectors
of KL-divergences using sets of 20 sequential empty room samples. From
the two sets of 100 empty room samples, this leads to 162 sets of
sequential samples. The initial choice of $\lambda$ was the same used
in Experiment A. The overall RMSE was calculated for each of these
sets of empty room samples. Two of the 30 person locations had a wide
variation in their range estimate depending on which set of empty room
samples was chosen. Figure \ref{fig:er_ecdf} shows the empirical CDF
of the final RMSE obtained using each of these sets of empty room
samples both with and without these two person locations. 

For the trials using all person locations, 12.3\% of the trials
resulted in an RMSE better than the 1.33\,ns achieved using all of the
empty room samples together. The overall RMSE for all of the trials
using 20 empty room samples is 4.19\,ns. This illustrates that, on
average, using a fewer number of empty room samples degrades
performance.

\subsection{False Positives}
Testing for false positives, or non-zero estimates of $k_*$ in empty
room samples, was also performed. The Baum-Welch algorithm was not
performed on these samples, that is, no updating of the HMM parameters
was done for re-estimation of $k_*$.

False positives were tested by randomly dividing the set of
empty-room samples into the known empty-room and possible point
sample sets. Due to the limited sample sizes for empty-room
samples, this random set division allows us to simulate how false
positive tests might perform using different sample sets that aren't
available. For each radio pair, the available samples were
divided evenly between the known empty-room sample set and the
possible point sample set. These two sets were used to find the
observation vector of KL divergences, which the HMM uses to estimate
$k_*$.

For each of the six transmitter/receiver pairs for each of the two
rooms, 1000 trials were performed using the random subset division
described for a total of 12,000 trials. Of these a total of 50 trials
resulted in false positives, that is, a $4.2\times 10^{-3}$ false
positive rate.  We note that over half of the false positives come
from a single transmitter/receiver pair in one of the rooms. Notably,
this pair had just 10 empty-room samples available for testing. This
is the fewest number of empty-room samples for any transmitter
receiver pair.

\subsection{Localization} \label{sec:localization_results}

Results for localization are given for both the forward method
described in Section \ref{sec:localization_methods} and the SLA
described described by Cheng Chang et al.\ \cite{Chang}. The forward
solving method is done in two ways, first using $\alpha_k$ where
$\alpha_k = P(X_k=1|\mathbf{O},\lambda)$ and second using only the
range estimates, $\hat{k}_*$, without the additional information of
the probability of being in a given state.

The SLA described by Cheng Chang et
al.\ only uses range estimates for localization. A summary of the
results of each localization method with its available information is
given in Tables \ref{tab:localrms} and \ref{tab:localmedian}. All
values are given in cm.

\begin{table}[htbp]
  \centering
  \caption{RMS Localization Error (cm)}
    \begin{tabular}{cccc}
    \toprule
          & \multicolumn{2}{c}{Forward} & SLA \\
    \midrule
          & All Info & Range Only & Range Only \\
    Rm 3325 & 36    & 155   & 165 \\
    Rm 1208 & 24    & 75    & 194 \\
    \bottomrule
    \end{tabular}%
  \label{tab:localrms}%
\end{table}%

\begin{table}[htbp]
  \centering
  \caption{Median Localization Error (cm)}
    \begin{tabular}{cccc}
    \toprule
          & \multicolumn{2}{c}{Forward} & SLA \\
    \midrule
          & All Info & Range Only & Range Only \\
    Rm 3325 & 16    & 67    & 159 \\
    Rm 1208 & 16    & 29    & 172 \\
    \bottomrule
    \end{tabular}%
  \label{tab:localmedian}%
\end{table}%

The forward solving method described here gives location estimates
that are significantly better than those from the SLA
described by Cheng Chang et al. Taking into account $\alpha_k$
rather than using $\hat{k}_*$ alone also improves the location
estimates for the forward solving method.  

\begin{figure}[h]
	\centerline{\epsfig{file=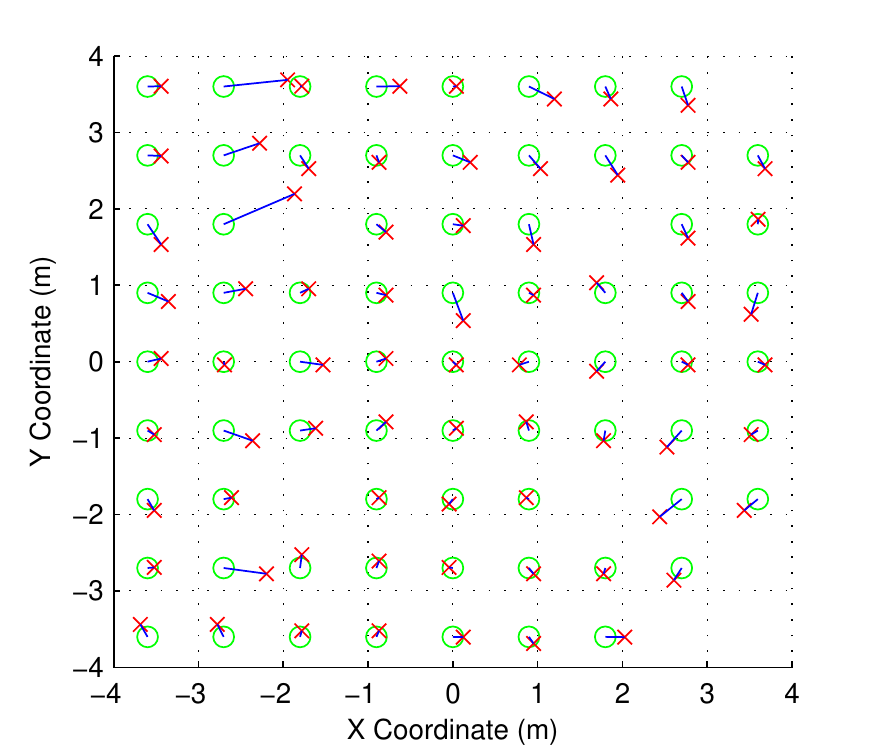} }
	\caption{MEB 3325 actual person positions (\textcolor{green}{O}) and localization estimates (\textcolor{red}{X}) using the forward solving method.}
	\label{fig:room1local}
\end{figure}

\begin{figure}[h]
	\centerline{\epsfig{file=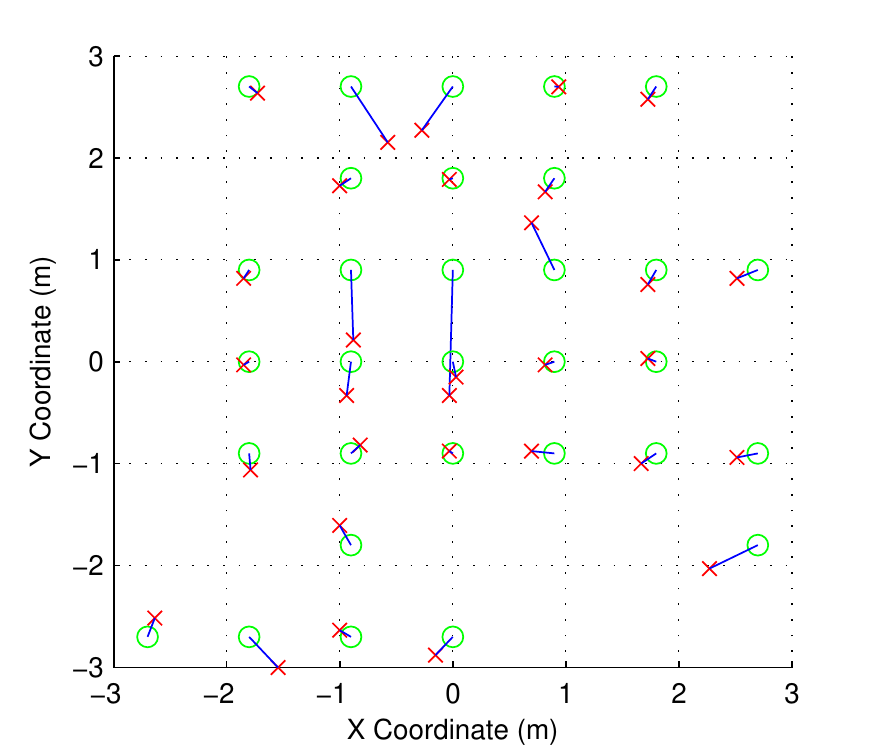} }
	\caption{MEB 3325 actual person positions (\textcolor{green}{O}) and localization estimates (\textcolor{red}{X}) using the forward solving method..}
	\label{fig:room2local}
\end{figure}

Figures \ref{fig:room1local} and \ref{fig:room2local} describe the
true person locations, as shown previously in Figure
\ref{fig:rooms1and2}, and the estimates for those locations using the
forward solving method with all available information.

\section{Conclusions} \label{sec:conclusions}
In this paper, we introduce and experimentally-verify a hidden Markov model-based 
algorithm for estimating the bistatic delay in an UWB impulse radar system. 
We show the proposed algorithm achieves a lower RMSE than first threshold crossing methods for highly
cluttered multipath environments. Applying the Baum-Welch algorithm
allows the proposed estimator to adapt its parameters to be best for the particular environment.  We show
the algorithm is robust to initialization parameters derived from a different environment.

Compared to using the first threshold crossing estimate of $\tau_*$,
our method reduces error by almost half. Since these estimates of the
person's bistatic delay are used directly in tracking algorithms, we
expect to similarly improve UWB-based localization performance.

The forward solving method described here for localization using the
probabilities $\alpha_k$ was very effective, achieving a median error of 18 cm.

One primary limitation of the algorithm as proposed is that it assumes only one person 
is causing changes to the CIR. To account for more people, future work must expand 
the HMM-based estimator to estimate a bistatic delay for each person in the environment.  Research
must determine what methods to use in the multiple person case, for example, if more states are
needed in the Markov model, or if joint estimation the number of people and their bistatic delays improves performance.

\bibliographystyle{IEEEtran}
\bibliography{IEEEabrv,wisnet_2012}

\begin{IEEEbiography}[\psfig{figure=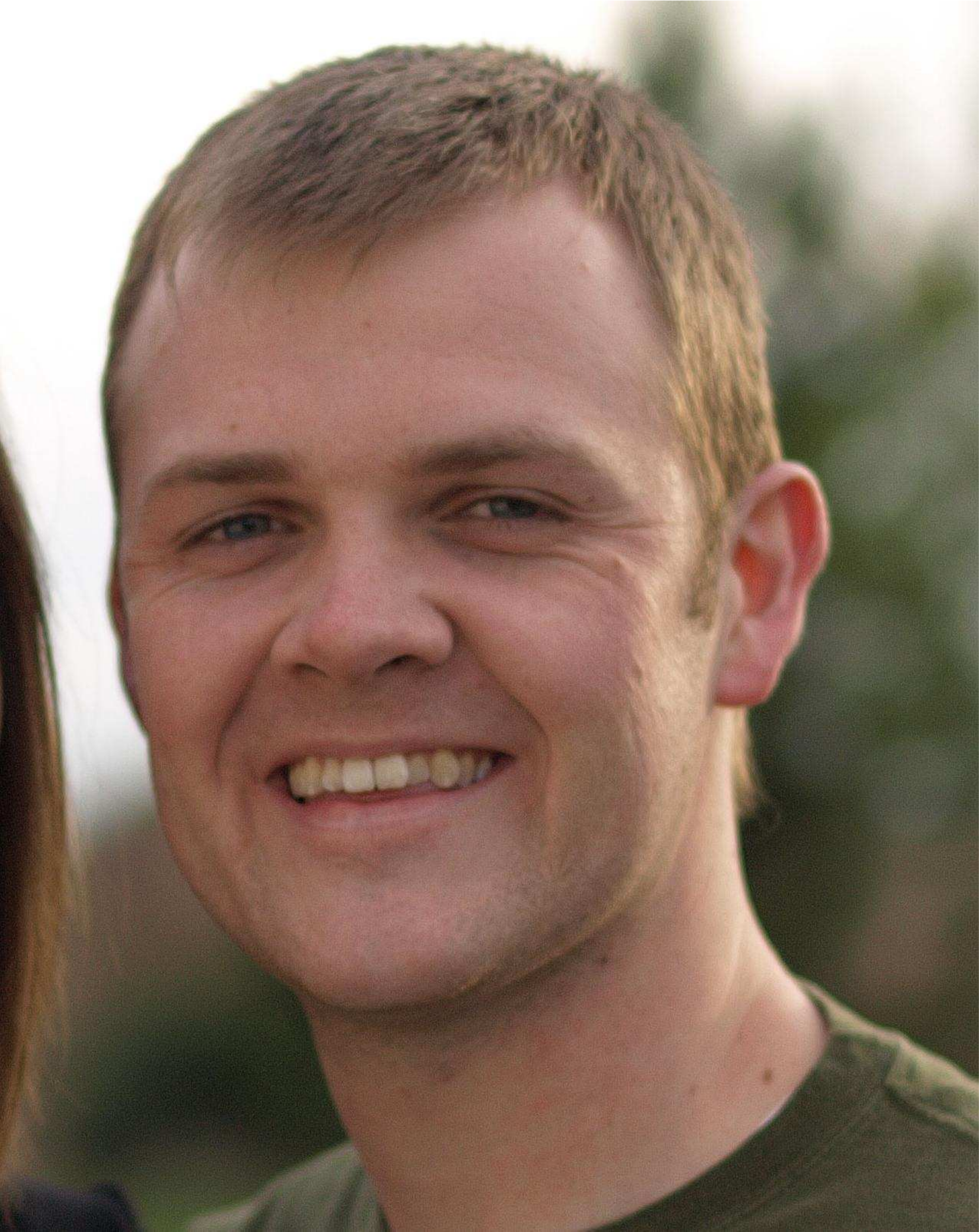,width=1.0in,height=1.25in}]{Merrick
  McCracken}
received the B.S. in Computer Engineering (2009) from Brigham Young
University - Idaho and the M.S. in Electrical and Computer Engineering
(2011) from the University of Utah. He is a recipient of of the
Science, Mathematics \& Research for Transformation Scholarship (2010)
through the US Department of Defense. He is currently working as a
Ph.D. student in the Sensing and Processing Across Networks (SPAN)
Lab at the University of Utah. 
\end{IEEEbiography}

\begin{IEEEbiography}[\psfig{figure=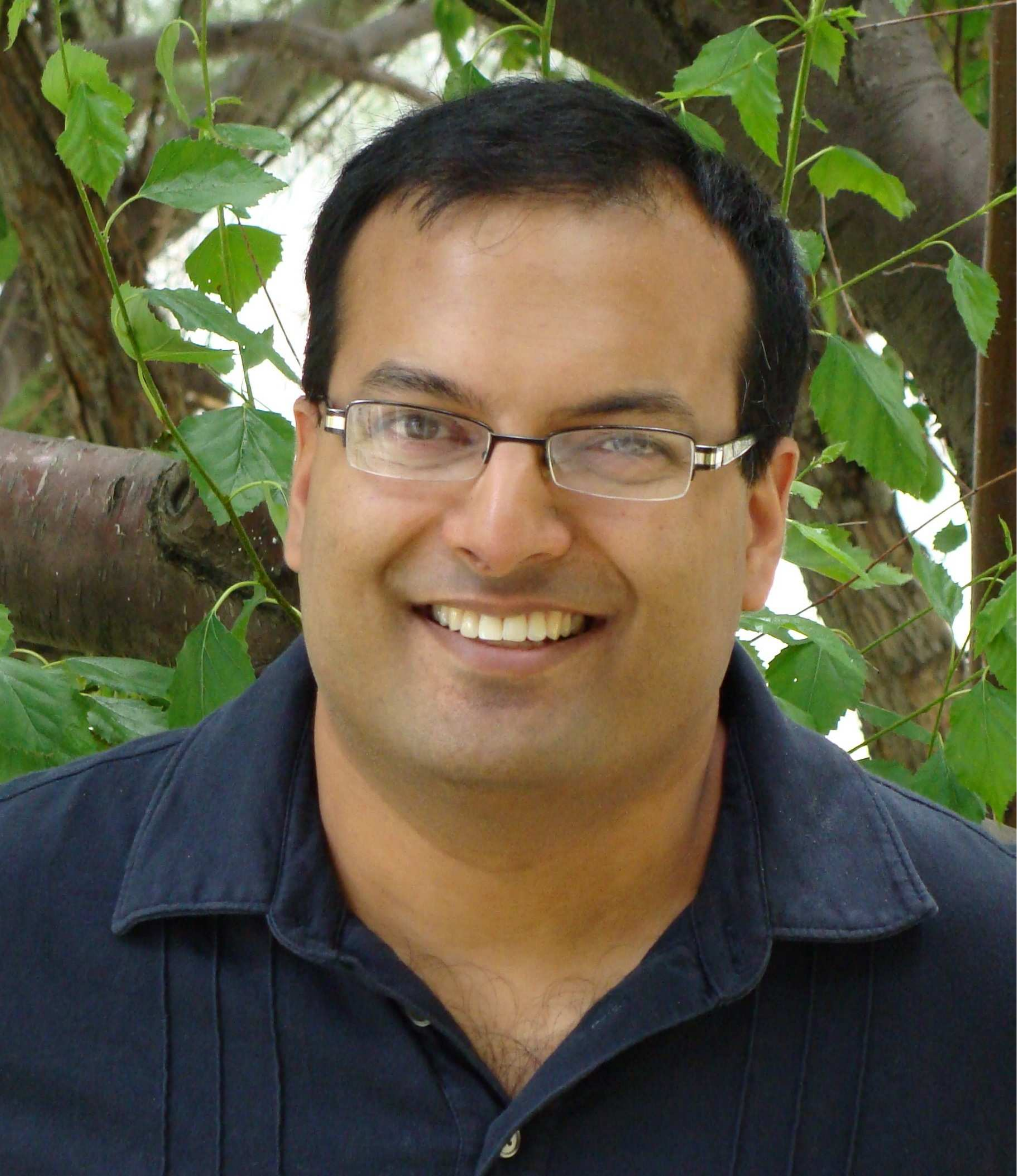,width=1.0in,height=1.25in}]{Neal
    Patwari}
  received the B.S. (1997) and M.S. (1999) degrees from Virginia Tech,
  and the Ph.D. from the University of Michigan, Ann Arbor (2005), all
  in Electrical Engineering. He was a research engineer in Motorola
  Labs, Florida, between 1999 and 2001.  Since 2006, he has been at
  the University of Utah, where he is an Associate Professor in the
  Department of Electrical and Computer Engineering, with an adjunct
  appointment in the School of Computing.  He directs the Sensing and
  Processing Across Networks (SPAN) Lab, which performs research at
  the intersection of statistical signal processing and wireless
  networking. Neal is the Director of Research at Xandem, a Salt Lake
  City-based technology company.  His research interests are in radio
  channel signal processing, in which radio channel measurements are
  used to benefit security, networking, and localization applications.
  He received the NSF CAREER Award in 2008, the 2009 IEEE Signal
  Processing Society Best Magazine Paper Award, and the 2011
  University of Utah Early Career Teaching Award. He is an associate
  editor of the IEEE Transactions on Mobile Computing.
\end{IEEEbiography}

\end{document}